\documentclass[reqno]{amsart}
\usepackage{graphicx}
\usepackage{epsf}
\usepackage{color}
\def\be{\begin{equation}}
\def\ee{\end{equation}}
\def\ba{\begin{array}}
\def\ea{\end{array}}
\def\bea{\begin{eqnarray}}
\def\eea{\end{eqnarray}}

\def\noi{\noindent}

\begin{document}

%

\title[Scientific philosophies and Atwood's machine]{Spreading scientific philosophies with instruments: the case of Atwood's machine}

\author{S. Esposito}
\address{{\it S. Esposito}: I.N.F.N. Sezione di
Napoli, Complesso Universitario di M. S. Angelo, Via Cinthia,
80126 Napoli ({\rm Salvatore.Esposito@na.infn.it})}%
\author{E. Schettino}
\address{{\it E. Schettino}: Dipartimento di Scienze Fisiche,
Universit\`a di Napoli ``Federico II'', Complesso Universitario di
M. S. Angelo, Via Cinthia,
80126 Napoli ({\rm Edvige.Schettino@unina.it})}%

\begin{abstract}
We study how the paradigm of Newton's science, based on the organization of scientific knowledge as a series of mathematical laws, was definitively accepted in science courses - in the last decades of the XVIII century, in England as well as in the Continent - by means of the ``universal'' dynamical machine invented by George Atwood in late 1770s just for this purpose. 

The spreading of such machine, occurred well before the appearance of Atwood's treatise where he described the novel machine and the experiments to be performed with it, is a quite interesting historical case, which we consider in some detail. In particular, we focus on the ``improvement'' introduced by the Italian Giuseppe Saverio Poli and the subsequent ``simplifications'' of the machine, underlying the ongoing change of perspective after the definitive success of Newtonianism. 

The case studied here allows to recognize the relevant role played by a properly devised instrument in the acceptance of a new paradigm by {\it non-erudite} scholars, in addition to the traditional ways involving erudite scientists, and thus the complementary role of {\it machine philosophy} with respect to mathematical, philosophical or even physical reasoning.
\end{abstract}

\maketitle



\section{Introduction}

\noi The XVIII century was characterized by a widespread interest for natural philosophy, driven by public lecture courses and the publication of popular texts in addition to the traditional forms of transmission of knowledge. The scientific societies and academies created during the scientific revolution played an increasingly larger role than universities, thus becoming the cornerstone of organized science \cite{soc}.
In England, the presidency of Sir Isaac Newton to the Royal Society from 1703 to his death in 1727 obviously led British science to be dominated by Newtonian ideas and experiments, and the structure itself of Newtonian physics, that is scientific knowledge organized as a series of mathematical laws, became the model for all sciences. However, due to this use of mathematics, very few people - in England and abroad - were able to understand Newton's ideas, so that an army of popularizers, lecturers and textbook writers was required in order to widespread Newtonianism: this occurred precisely in the early XVIII century, continuing for about half a century.

The rapid dissemination of Newton's science came first via the members of the Royal Society, both British and Continental \cite{Feingold}, and an important role was played by Huguenots \cite{ugo}, i.e. French Protestants who had been expelled from France as early as 1685. In this respect, a key contributor to the success of Newtonian science was John Theophilus Desaguliers, whose work of dissemination extended well beyond his lectures and publications. Indeed, for example, in 1715 Desaguliers also performed experiments on colours before members of the French Acad\'emie Royale des Sciences, thus paving the way to a wider recognition of the validity of NewtonÕs theories in France, and much the same took place before diplomatic audiences from Spain, Sicily, Venice and Russia \cite{Stewart}. The Dutch Willem Jacob 's Gravesande was present to the demonstration of 1715, and later became a decisive spokesman of Newtonian science in the Netherlands, a country which already played in the XVII century a significant role in the advancement of the sciences, including Isaac Beeckman's mechanical philosophy and Christiaan Huygens' work on the calculus
and in astronomy \cite{Jacob}. The Netherlands became the first Continental country to adopt Newtonianism, with Amsterdam becoming the center of European publishing, thanks, again, to French exiles, while in France the struggle against Cartesianism lasted for some time, the Newtonianism attracting mainly people dissatisfied with French society and the Catholic Church, like Voltaire. A similar, or even deeper struggle, was fighted in the Germanic world, where Newton's science contended against the prestige of Leibniz's philosophy, this resulting in the long-lasting Newtonian-Wolffian controversy \cite{Wolff}: in the XVIII century, Germany was always hesitant about the mechanization of nature, and eventually Newton's mechanics was incorporated into the dominant German academic philosophy. Instead, Newton's science filtered quite early into different parts of Italy, notwithstanding the Inquisition looked with suspicion on these ideas, not least because they came from a Protestant country.

The traditional way to disseminate Newtonian natural philosophy was, of course, through academic courses: before the XVIII century, indeed, science courses were taught almost exclusively through formal lectures. Thus, in the first instance, scientists, mathematicians and philosophers of Holland, France, Germany and Italy read Newton's books and taught the ideas to their students. The structure of courses, however, began to change in the first decades of the XVIII century, when physical demonstrations were added to academic lectures, and popular lectures were introduced in response to the growing demand for science. For example, Pierre Polini\`ere was among the first individuals to provide demonstrations of physical principles in the classroom: he presented experimental demonstrations of his own devising before students at the University of Paris \cite{Hanna}, his lectures proving so popular that in 1722 he presented a series of experiments before Louis XV, King of France. In particular, in physics, such a change coincided with the institutionalization of the course in Experimental Physics \cite{Schet}.

Now, still in the XVIII century, what we now could call ``research centres'' of the newborn physics, were not necessarily distinct from the places devoted to the transmission of knowledge, i.e. schools, colleges, etc. The success of a theory also passed through its acceptance by ``students'', i.e. scholars not yet acquainted with it, who then contributed actively in the dissemination of novel ideas. In other words, if the first acceptance of a given doctrine was disputed at a rather ``high" level between leading scientists and philosophers (as a reminiscence of the philosophical disputations of earlier centuries), the definitive one was played in lecture halls before non-experts. An illuminating example of this mechanism is just the general achievement of Newton's mechanics, expressed in his {\it Principia} \cite{Principia}. While its acknowledgement by leading scholars is well documented in the literature \cite{NewtonSuccess}, the same is unfortunately not true for its thorough acceptance by ``students'':\footnote{See, however, the beautiful review by Schaffer on the role of  ``machine philosophy'' in the XVIII century's England \cite{Schaffer}.} in this paper, we will focus just on this last historical case.

If treatises of natural philosophy played a major role in the ``high" level debates, the writing and the circulation of practical textbooks for the courses in Experimental Physics (or similar other ones) became essential for the dissemination and acceptance of the novel ideas among ``students''. In the following Section we then review this issue. Far from being exhaustive, we focus just on the most known scholars and their textbooks in England and in the Continent, by concentrating on the period immediately following that influenced directly by Newton. Instead, in Section \ref{s3} we enter into the heart of the subject matter hereof, by showing how the invention of a ``universal'' dynamical machine by George Atwood, described along with the experiments to be performed with it in an accompanying treatise, contributed substantially to the final, general achievement of Newton's mechanics. The invaluable work of Jean-Hyacinthe de Magellan and especially that, much less known, of Giuseppe Saverio Poli in the rapid spreading of Atwood's machine outside England is considered in Section \ref{s4}, where the subsequent key changes and ``simplifications'' underlying the ongoing change of perspective are pointed out as well. Finally, in the concluding section we summarize what discussed in order to provide a rapid overview of the historical case studied.

\

\

\section{Newtonian textbooks of natural philosophy in the post-Newtonian age}
\label{s2}

\noi Newtonian doctrines spread in the first quarter of the XVIII century thanks to elementary or complex textbooks by English authors such as J.T. Desaguliers, or the Dutch ones W.J. 's Gravesande and Pieter van Musschenbroek, and the Frenchman Jean Antoine Nollet. The complexity of Newtonian thought expressed itself in several different ways, and in the following we will give a glimpse of such diversity. 

\subsection{Newton's experimenter successor}

The first English disciples of Newton tried to give an experimental basis especially to ideas that he had not sufficiently formalized (such as that of ether and that of interparticle forces), so that the results obtained by Newton by means of a complex relationship between inductive and deductive reasoning became truths demonstrable directly with experiments.

The Newtonian method underwent a profound remodeling with Desaguliers, a member of the Christ Church College who held popular Newtonian lectures at Oxford, where he had much success in his activity \cite{Desag}. After moving to London in 1710, he was welcomed into Newtonian circles of the capital and was then (1714) elected fellow of the Royal Society, later becoming the {\it curator} of the Royal Society, a position that was earlier held by Francis Hauksbee. Desaguliers published many works on the {\it Philosophical Transactions} about experiments on heat, mechanics and electricity. In particular, he correctly realized that the physical quantities then used for describing the motion of a body, i.e. the {\it momentum} for the Newtonians and the {\it vis viva} for the Leibnizians, were different concepts, and that the Newtonian one was to be preferred because better supported by experiments.

As curator of the Royal Society, Desaguliers performed experiments that were repetitions of those already described by Newton, although there were more complex experiences such as those aimed at finding a similarity between the electric force and the force of cohesion \cite{AtomPowers}. However, Desaguliers' reputation as a scientist was sealed (apart by his three awards from the Royal Society) by publication of a two volume work about {\it A Course of Experimental Philosophy} \cite{DesagCourse}, whose first volume was published in 1734, while the second volume's publication came 10 years later in 1744. The first volume deals with mechanics, with an explanation of the basics of Newtonian science, while the second volume contains  material oriented toward practical application of scientific findings. Desaguliers contributed significantly to the wide spread of Newtonian-oriented textbooks also by translating Edm\'e Mariotte's {\it Trait\'e du mouvement des eaux et des autres corps fluides} \cite{Mariotte}, as well as 's Gravesande's Latin treatise on {\it Physices elementa mathematica, experimentis confirmata} \cite{Gravesande}. In addition, he also wrote several texts, among which we recall {\it The Newtonian system of the world, the best model of government: An allegorical poem} \cite{DesagPoem} and {\it A dissertation concerning electricity} \cite{DesagDiss}, which received a prize awarded by the French Acad\'emie de Bordeaux.

\subsection{Instrumental philosophy in the Netherlands}

The introduction of Newtonian science in the Netherlands and, more in general, in the Continent came through 's Gravesande \cite{sGrave}. Educated at the University of Leiden, in 1714 he had the opportunity to be part of the delegation sent to England by the Dutch States General to congratulate King George I on his accession to the throne, and just during his (one year) stay in London he attended sessions of the Royal Society, later being elected to membership, and made acquaintance of Newton and, especially, Desaguliers. On his return to Holland, 's Gravesande was appointed professor of astronomy and mathematics at the University of Leiden, and in 1720 published the first of the two volumes on {\it Physices Elementa Mathematica, experimentis confirmata} \cite{Gravesande}, already mentioned above.

Written in Latin, this first Newtonian-oriented textbook on natural philosophy was of course accessible to all educated readers in Europe, but the English translation performed by Desaguliers in the same year of its publication certainly contributed to its rollout. In the discussion on the infinite divisibility of matter and on small real particles of which it was composed, Newton's hypothesis on particle different shape and size was not discussed by 's Gravesande, but simply ignored: his position in favor of Leibniz's theses had revived the debate on the {\it vis viva}. Instead, 's Gravesande agreed with Newton about the role played by short-ranged forces of attraction and repulsion acting on the fundamental particles, the rising of liquids by capillary action being considered as an example supporting the Newtonian theory (such phenomena were treated as {\it Queries} in Newton's {\it Opticks} \cite{Opticks}). 's Gravesande's treatise displayed a complete adhesion to Newtonian approach, not only about the structure of matter and short- and long-ranged forces, but also on questions of mechanics and astronomy. However, he took no position about ether; rather, on this topic,  's Gravesande showed to follow the ideas of his teacher Herman Boerhaave, three chapters being devoted to the discussion on the fire and its nature \cite{AtomPowers}.

In the third decade of the XVIII century, Dutch science enjoyed a period of great splendor with s' Gravesande's successor in Leiden, Pieter van Musschenbroek \cite{Muss}, who followed the same approach introduced by his predecessor. The lectures delivered by Musschenbroek, indeed, maintained the excellent reputation that Boerhaave and 's Gravesande earned to the University of Leyden, where students interested to experimentation flocked from all over Europe. The Musschenbroeks were scientific instruments' makers since the seventeenth century, Joost (1614-1693) being the founder. His sons Samuel and Johan continued such activity, specializing themselves in the construction of microscopes, telescopes and air pumps, all of them being typical instruments used at that time. Pieter and Jan were the sons of Johan, and both had an excellent education, his teacher being Boerhaave; however, while Pieter decided to pursue his academic studies, Jan chose to continue the family business. 

Pieter van Musschenbroek began his academic career in 1719 in Duisburg, where he taught mathematics and philosophy, and then, after a temporary moving to Utrecht, finally (in 1739) came to Leiden, where he succeeded to 's Gravesande. His lectures were characterized by a systematic use of experimental devices, many of them being constructed by his brother Jan. Skilfully designed models, illustrating the use of those machines and the experiments that could be run with them, supplemented the textbooks written by Pieter, among which we find the {\it Elementa physicae} of 1734 \cite{Elementa} and the {\it Institutiones physicae} of 1748 \cite{Insti}. The large diffusion of these textbooks also increased the demand for instruments constructed by Jan: both universities and private science amateurs wanted to buy them for their demonstrations. The demand increased to such an extent that the manufacturers of scientific instruments began to reproduce the same models of Jan: it 's well known, for example, that the instruments made by George Adams in the cabinet of George III of England were inspired just by those models.
The fame of Pieter van Musschenbroek spread rapidly throughout Europe, and his works were translated into German, French and English, in addition to further editions of the {\it Elementa physicae}, including the Neapolitan one of 1745.

\subsection{French Newtonianism}

The Newtonian science was introduced in France through Dutch scholars. Newton's {\it Opticks}, considered as a book of experimental physics, was well and quite soon received (a translation into French was published as early as 1720 \cite{Optfra}), given the pronounced experimentalism of the Acad\'emie des Sciences, while the diffusion of the {\it Principia} encountered some more difficulty. An initial English corpuscular approach, followed by the experimental Newtonianism in France, gave then way to the typically Dutch fluidic approach, this change being facilitated by some Newton's early reflections on the ether, published only in 1744 \cite{alllight}. This transition from discontinuous to continuous proved very heuristic, especially in the studies about electricity. 

Charles-Francois de Cisternai Dufay realized, after many experimental observations, that two different kinds of electricity - the vitreous and the resinous ones - existed: together with his collaborator Jean Antoine Nollet, he proposed to explain electrical phenomena starting from the existence of two opposite fluid currents, gave off by electrified bodies. The success of the fluidic approach was accompanied by the abandonment of the idea on the homogeneity of matter, and fire, ether and electric fluids were conceived as substances different from ordinary matter. 

Like many other young scholars from all over Europe, Nollet \cite{Nollet} was attracted by the fame of s' Gravesande and moved to Leyden to follow his lectures. On his return from Holland, after a visit to London, where he was admitted into the Royal Society, in 1735 he succeeded to Pierre Polimi\'ere who taught at the College of the University of Paris, and in 1740 became a member of the Acad\'emie des Sciences. Nollet's contribution to the spread of experimental physics was quite remarkable and his treatises, among which we mention the {\it Lecons de physique experimentale} \cite{Lecons}, published in six volumes between 1743 and 1748 and often reprinted, enjoyed enormous popularity. This textbook offered many experiments on electricity, and thanks to a very substantial contribution by Voltaire, Nollet could realize many instruments, about 350, with which he performed experiments both during public lectures and in his physics courses.

It was, instead, quite difficult for the Newtonian astronomy, based on the concept of action at a distance, to be introduced in France, given the wide spread of the Cartesian theory of vortices. Through the reading that the Dutch scholars had made of Newton's works, however, Voltaire reinterpreted the {\it Principia} along empiricist lines \cite{Voltaire}, thus reconciling it with the experimentalism of the French Acad\'emie des Sciences. Still the conflict between forces at a distance and theory of vortices remained to be resolved, and this happened with a great theoretical debate, which concerned the compatibility of the theory of vortices with the Kepler's laws and with the theory on the shape of the Earth. Indeed, vortices proved to be incompatible with Kepler's laws, and the contrast between the Cartesian idea of the Earth lengthened at the poles - as early measurements performed by the astronomer G. Cassini seemed to show - against the idea of Newton on the flattening at the poles due to the proper motion of the planet, found a definitive solution with the measurements of Earth's curvature at the equator and at the Arctic Circle \cite{curvature}. This theoretical debate did not find room in empiricist oriented textbooks, but it should be stressed that such studies led to the development of rational mechanics, carried on by mathematics scholars such as P.L.M. de Maupertuis, A.C. Clairaut, L. Euler, J.B. d'Alembert and the Bernoullis, who were poorly interested in experimenting.

\subsection{Experimental science in Italy}

Physics textbooks that privileged an empirical approach were inspired, in Italy,  by the works by 's Gravesande and Musschenbroek, and the University of Naples, in particular, was among the first to introduce in 1734 the teaching of experimental physics. The sponsor of the institution of this chair was the Grand Chaplain Celestino Galiani, a great supporter of Newtonian ideas, who entrusted the chair to Giuseppe Orlandi \cite{Schet}. The intent to pursue a teaching based on the inductive method, as practiced in the Dutch universities for over ten years, was clear already from a letter by Galiani to the Grand Chaplain: ``In the beginning, a preface will be present in which I will show how the true way of philosophizing is nothing but by means of experiments; and compare the state of philosophy among scholastics lasted for many centuries with the present, started by so many talented scholars who used to look at the book of Nature and study the characters with which it is written, which are observation, experiments and geometry'' \cite{Palladino}.

It is evident, however, that in order to perform experiments, as claimed by the followers of empiricism, several machines were required, i.e. instruments by means of which the laws of physics could be inductively derived. For this reason, Orlandi adopted the {\it Elementa physicae} by Musschenbroek, where the necessary devices were skilfully depicted, but added to it a long appendix on astronomy - {\it De rebus coelestibus tractatus}. The first Naples edition of this texbook appeared in 1745 \cite{Petro}, with a long introduction written by Antonio Genovesi under the meaningful title {\it Disputatio physico-historica de rerum corporearum origine et constitutione}, where the illustrious Neapolitan follower of the Enlightenment attempted to give a first, brief but insightful history of physics from ancient times to XVIII century \cite{Torrini}.

Giovanni Maria della Torre, entrusted with the teaching of experimental physics in the Royal Archigymnasium in Naples, was the author of a very successful textbook published between 1748 and 1749 in two volumes \cite{DellaTorre}. The first volume - {\it Scienza della Natura generale} - was devoted to general physics, that is Della Torre expounded about ``matter, extent, strength, mobility and motion'', while in the second one - {\it Scienza della Natura particolare} - topics related to Earth science were discussed, that is about ``Earth's shape and size, internal structure, surface and atmosphere''. The della Torre's textbook was different, especially in the second volume, from the most authoritative ones by Musschenbroek and 's Gravesande, where biology and geology were omitted, as well as almost all of chemistry and meteorology. The historical and critical notes about ``ancient conceptions and their links with recent results'', though always present in Newtonian oriented textbooks, were so many in the della Torre's treatise that raised strong debates \cite{Torrini}. In this book, as well as in any contemporary textbook, the use of mathematics was always theorized, but scientific theories were explained without resorting to analytic calculations. This, however, did not mean that theories were not addressed in a rigorous way: for example, in Section IV of the second volume, detailed theorems were present in order to justify the laws of optics and catoptrics. As in the {\it Lecons} by Nollet, a large part was devoted to the description of instruments, especially microscopes, with fully spherical lenses (invented by della Torre himself), which limited the problem of spherical aberration. della Torre acquired such a skill in experimenting with microscopes that was famous throughout Europe: J.J. de Lalande, who traveled in Italy between 1765 and 1766, had him in high regard, and quoted him often in his book {\it Voyage en Italie} \cite{Lalande}, and the same applies to Nollet, who had the opportunity to meet him during his stay in Naples in 1750. 

Later in the XVIII century, starting from 1770s, a new generation of enlightened scientists, who were educated at the University of Padua, where they had very strong links with scientific circles in England,
began to impose in the academies of the Kingdom of Naples \cite{Borrelli2}. Giuseppe Saverio Poli was one among these who, having had the opportunity to stay in Cambridge and meet George Atwood \cite{PoliToscano}, was the first to publish a new, Newtonian oriented textbook \cite{PoliElementi} where the novel machine invented by the English scholar was illustrated with a choice of experiments. 

Now, however, one step back is needed in order to fully understand the change of perspective that occurred in the last quarter of the XVIII century, and that further influenced the subsequent transmission to ``students'' of physics results.

\

\

\section{Atwood and the general achievement of Newton's mechanics}
\label{s3}

\noi  The Atwood's machine changed the way of propagating Newtonian mechanics, and had also a prominent role in the definitive success of Newtonianism, though this fact is not always correctly realized or even considered. In the following we will focus just on this issue, while in the subsequent section we will show how Poli and others contributed to the widespread use of Atwood's machine (and, in particular, to a specific use of it) in the Continent.

\subsection{Popular lecturer in Cambridge}

Biographical notes on Rev. George Atwood are scarse, and mainly
related to his record in the {\it Alumni Cantabrigienses} list \cite{Alumni} or
other similar records in England (see \cite{Bios}). He was the
first of three sons (with James and Thomas) by Isabella Sells of
Inglesham, Wiltshire, and Thomas Atwood, the curate of the parish
of St. Clement Danes, Westminster, where George was baptized on 15
October 1745. After attended the Westminster School, starting in
1759, as a king's scholar, he entered Trinity College in Cambridge
on 5 June 1765 as a pensioner (i.e. he paid for his own keep in
College), and was then elected to a scholarship on Lent 1766,
being also awarded with the Members' prize in the same year. He
graduated (as third Wrangler) with a B.A. in 1769, being first
Smith's prizeman in the same year, and received his M.A. in 1772.
Meanwhile, in October 1770, Atwood became a Fellow of Trinity
College and taught there, also becoming a tutor in 1773. His
lectures in the observatory over the Great Gate of Trinity College
were well attended and received because of their delivery and
their experimental demonstrations. He published descriptions of
his demonstrations in 1776 \cite{Description}, the year he was
elected Fellow of the Royal Society: they consisted of simple
experiments to illustrate mechanics, hydrostatics, electricity,
magnetism and optics. One of the many students who attended
Atwood's popular lectures was William Pitt, who later (in 1783)
achieved the high office of British Prime Minister. In 1784 Atwood
was then hired to a major post in the customs office as part of
Pitt's campaign for administrative rationalization: he ``rendered
important financial services to Pitt, who bestowed upon him a
sinecure office, as one of the patent searchers of the Customs,
with a salary of \pounds 500'' \cite{Alumni}.

Atwood is now best known for a textbook on Newtonian mechanics,
{\it A Treatise on the Rectilinear Motion and Rotation of Bodies}
\cite{Treatise}, published in 1784, where he also describes in
detail a machine, now known as Atwood's machine (see below). In
the same year he also published a second work, {\it An Analysis of
a Course of Lectures on the Principles of Natural Philosophy}
\cite{Analysis}, which is an expanded version of his Cambridge
course which he had first given detail in 1776.

Most of other Atwood's published works \cite{Biblio} consists of
the mathematical analysis of practical problems, including a {\it
review} for Pitt in which he analyzed the cost of bread and
attempted to rationalize the standards for it. A particular
mention merit his works on the stability of ships \cite{17961798},
where he extended the theories of Euler, Bougier and others to
account for the stability of floating bodies with large angle of
roll, and for which, in 1796, he was awarded the Copley Medal of
the Royal Society. Finally, he also wrote on the construction of
arches \cite{arches} and on the design of a new iron London Bridge
over the Thames at Blackfriars.

Atwood died ``unmarried'' \cite{Alumni} on July 1807 and buried at
St. Margaret's, Westminster, where his brother Thomas had
succeeded his father as curate.

\subsection{A novel machine}

The name of Atwood is almost entirely related to the dynamical
machine he invented between 1776 and 1779: the story of the
development of this machine and the spreading of its existence
throughout Europe is, by itself, worth-mentioning, and will be
considered in some detail here and in the next Section. Its scope,
according to textbooks appeared since the end of 1700s till recent
times, would be just that of conducting experiments proving the
laws of rectilinear (and rotational) motion of bodies, with
particular reference to motions ruled by gravity. As we will see,
such a (reductive) purpose of Atwood's machine was established
only {\it after} the general achievement of Newton's mechanics,
that is from the end of the XVIII century onward, but in the
Cambridge of 1770s and 1780s it served just to fulfill
definitively such achievement before young students and scholars
who {\it later} disseminated the Newtonian paradigm.

\begin{figure}
\includegraphics[width=60mm]{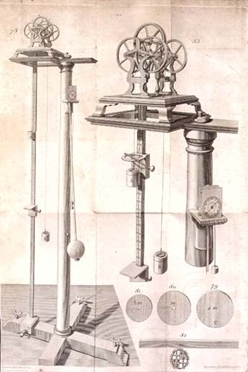}
\caption{The plate in Atwood's {\it Treatise} of 1784 \cite{Treatise} with the illustration 
of the novel machine built by Adams.}
\label{matwood}
\end{figure}

The machine is described in Atwood's {\it Treatise} of 1784
\cite{Treatise}, but its diffusion outside England dates back to
the end of 1770s (see next Section). As well known, it consists
(see Fig. \ref{matwood})
just of two balanced cylinders linked by a silk cord suspended
over a pulley, where additional weights can be attached (and
removed) to either cylinder in order to provide a net (or zero)
force acting on the system.

It is quite instructive to follow closely how Atwood deviced his
ingenious machine, starting from his first considerations about
the classical problem of experimenting on the free fall of bodies.
\begin{quote}
The most obvious method would be to observe the actual descent of
a heavy body, as it falls toward the earth by its natural gravity:
but in this case it is manifest, that on account of the great
velocity generated in few seconds of time, the height from which
the observed body falls must be considerable. [...] \\
If to remedy this inconvenience bodies be caused to descend along
inclined planes, according to the experiments of the celebrated
author [Galileo] of this theory, by varying the proportion of the
plane's height to their lengths, the force of the acceleration may
be diminished in any ratio, so that the descending bodies shall
move sufficiently slow to allow of the times of motion from rest
being accurately observed; and the effects of the air's resistance
to bodies moving with these small velocities will be absolutely
insensible: the principal difficulty however which here occurs,
arises from the rotation of the descending bodies, which cannot be
prevented without increasing their friction far beyond what the
experiment will allow of (\cite{Treatise}, pp.295-6).
\end{quote}
Notably, here the ``principal difficulty'' is with what theory
already knows (as explained in the first half of the book), namely
that the acceleration of a rotating body while descending along an
inclined plane is reduced with respect to the case where rotation
does not occur, and thus gravity is the sole cause of such effect.
While this is not accounted for in many  modern textbooks, the
problem envisaged  by Atwood is not the consideration of the
effect of rotation (he knew the correct reduction factors: $5/7$
for a sphere and $2/3$ for a cylinder), but simply the
superposition of {\it two} effects (rotation and gravity) that
could cause - in Atwood's mind - not a univocal acceptance of the
Newtonian paradigm. This attitude is
confirmed one page after in the {\it Treatise}, where other,
different problems with the inclined plane are envisaged.
\begin{quote}
There are no means separating the mass moved from the moving
force; we cannot therefore apply different forces to move the same
quantity of matter on a given plane, or the same force to
different quantities of matter. Moreover, the accelerating force
being constant and inseparable from the body moved, its velocity
will be continually accelerated, so as to render the observation
of the velocity acquired at any given instant impossible
(\cite{Treatise}, pp.298).\footnote{Note that, according to
Atwood's terminology, ``moving force'' is for force,
``accelerating force'' for acceleration and ``quantity of matter''
for mass.}
\end{quote}
The first problem is that, with a given inclined plane (i.e.
height to length ratio) it is not possible to study the dependence
of the force acting upon the body on its mass, thus proving
Newton's second law of motion. The second, more involved
experimental problem is instead the impossibility to measure the
velocity of the body at any desired instant of time, since it
continually changes, and thus the impossibility to prove the time
law for the velocity in the uniformly accelerated motion.

What is, then, Atwood's aim? Surprisingly enough, he is aware that
Newton's second law will be generally confirmed only when
considering the effect of {\it variable} forces, but the
inadequacy of the experimental conditions of the epoch forced him
to turn to constant forces (as in the free fall or in the motion
along an inclined plane). This, however, does not mean for him to
make any other concession to the ease of the experiment (again, as
is instead the case in the free fall or in the motion along an
inclined plane).
\begin{quote}
Although it might be difficult to reduce the effects of variable
forces on the motion of bodies to experimental test, yet the laws
observed during the motion of bodies acted on by constant forces,
admit of easy illustrations from matter of fact. But in order to
render experiments of this kind satisfactory, they should
comprehend the properties of the moving forces, the quantities of
matter moved, and the velocity acquired, as well as the spaces
described and times of description: which general properties of
uniformly accelerated motion are not so much considered in books
of mechanical experiments as the subject seems to demand
(\cite{Treatise}, p.294).
\end{quote}
The boundaries of the problem are, thus, well depicted: how is it
possible to device a series of experiments with a {\it single}
machine with which force, mass, velocity, distances and times are
dealt with? Without previous suggestions by other authors (which
is, instead, the case of Desaguliers, as mentioned above, who
simply improves Galilei's experiment on free falling bodies),
Atwood focuses his attention on pulleys, already worthy of
consideration by men of science since many centuries. But here the
change of perspective is crucial: a conventional device employed
in statics as a simple machine is transformed in Atwood's hands
into a carefully scaled {\it dynamically} device capable of being
subjected to mathematical analysis for the illustration of the
Newtonian paradigm. In this light, indeed, a consistent
interpretation can be given to a number of kinematics and dynamics
problems - just theoretical exercises, not directly related to his
machine - appearing in Atwood's {\it Treatise}, most of shich not
at all considered in subsequent textbooks.\footnote{To this
regard, it is quite illuminating what writes still in 1922 a
reviewer of Atwood's {\it Treatise}, evidently considered just as
a modern textbook: ``though written in 1784, some of its pages are
greatly superior to those in many of the textbooks now in use in
schools, and one is particularly impressed by the stress laid on
experimental verification of the various laws of mechanics, and by
the extreme care shown in the planning and execution of the
experiments proposed'' \cite{Hall}.}
\begin{quote}
In the instrument constructed to illustrate this subject
experimentally, $A$, $B$ represent two equal weights affixed to
the extremities of a very fine and flexible silk line: this line
is stretched over a wheel or fixed pulley $abcd$, moveable round
an horizontal axis: the two weights $A$, $B$ being precisely equal
and acting against each other, remain in equilibrio; and when the
least weight is superadded to either (setting aside the effects of
friction) it will preponderate (\cite{Treatise}, p.299).
\end{quote}
The general problems envisaged above (too short falling times - in
free fall or on an inclined plane - require very large distances
traveled by the falling body, upon which, however, acts a non
negligible friction force, the body reaching also too great
velocities near the end of its motion) disappear at once: a small
mass unbalance induces small accelerations and velocities, thus
preventing unnecessary large falling height and, then, neglecting
the action of air resistance. But Atwood's aim is a {\it precise}
confirmation of Newton's laws, and if the realization of
accurately balanced weights, as well as small while removable
additional weights, is leaved to the art of skilled craftsmen, the
possible friction developed by the axle of the pulley merits
appropriate considerations.
\begin{quote}
When the axle is horizontal it is absolutely necessary that it
should be supported on friction wheels, which greatly diminish, or
altogether prevent the loss of motion which would be caused by the
friction of the axle, if it revolved on an immoveable surface
(\cite{Treatise}, p.338).
\end{quote}
Such an apparently tricky mechanism, introduced earlier probably by the
French clockmaker Henry Sully \cite{Sully}, was installed by
Atwood on the top of the cylindrical column of his machine, upon
which a weight operated clock is as well mounted in order to
perform time measurements, while a ruler with ``a scale about $64$
inches in length graduated into inches and tenths of an inch'' is
added on another vertical column to allow measurements of the
distances covered by the bodies suspended on the string. The
assemblage with the wheels is, interestingly enough, a removable
piece of the machine: Atwood was, indeed, creating a device
capable of studying rectilinear motions as well as rotation of
bodies, to which almost half of his {\it Treatise} is devoted. The
recognition of this important second part of mechanics will be
lost in subsequent descriptions of Atwood's machine in textbooks -
and further copies of the machine itself will no longer show the
{\it removable} assemblage with the friction wheels - but it
should not fall into oblivion the fact that Atwood was originally
aimed to conceive a ``universal'' machine suited for studying both
rectilinear and rotational motion. This is clearly testified by a
number of interesting and intriguing experiments, proposed in his
{\it Treatise}, to be realized with the aid of additional
accessories sketched in figures 84-88 of the book \cite{Treatise}.
However, while the main part of the machine ``was executed with
great mechanical skill, partly by Mr. L. Martin, and partly Mr. G.
Adams, mathematical instrument makers in London''
(\cite{Treatise}, p.337), such accessories, especially the
principal one drawn in figure 84 of \cite{Treatise}, were never
realized, probably for the change of working interests; see the
biographical notes above. Thus, Atwood's machine was
subsequently associated only to the studies on rectilinear motion
under the action of a constant force.

\subsection{Displaying the Newtonian paradigm}

How did Atwood obtain a satisfactory illustration of the Newtonian
paradigm with his machine? The first step is, obviously, to
reproduce the Galilei's law of proportionality between the
traveled distances $s$ and the square of the elapsed time $t$ in
the uniformly accelerated motion: $s \propto t^2$. Under the
``action of the constant force $m$'' ($m$, here the unbalanced net
moving force, corresponds to the weight of $1/4$ oz of matter; it
is considered as a standard quantity in Atwood's experiments),
\begin{quote}
if the times of motion be 1 second, 2 seconds, and 3 seconds, the
spaces described from rest by the descending weight $A$ in those
times will be 3 inches, $3 \times 4 =12$ inches, and $3 \times 9 =
27$ inches respectively; the spaces described from rest being in a
duplicate ratio of the times of motion (\cite{Treatise}, p. 318).
\end{quote}
The second step is to study the dependence of $s$ on the
acceleration $a$ of the descending body, here obtained - in units
of the free fall value $g$ - from the ratio of the unbalanced mass
to the total mass of the bodies on the pulley (in modern terms,
$a/g= \Delta m / m_{\rm tot}$):
\begin{quote}
It appears from these experiments, that when the times are the
same, the spaces described from rest are as the accelerating force
(\cite{Treatise}, p. 320)
\end{quote}
(recall that ``accelerating force'' means just acceleration). That
is to say, $s \propto a$. What else about the equation of motion
regarding the distance traveled? {\it Obviously}, that $a \propto
1/t^2$:
\begin{quote}
The latter part of the experiment shews, that if the space
described remains the same, while the time description is
diminished, the force of acceleration must be increased in a
duplicate proportion of the times' diminution (\cite{Treatise}, p.
321).
\end{quote}
From these three experiments, then, the full time law that, in
modern terms, we write as $s = \frac{1}{2} a t^2$, is completely
derived.\footnote{According to the introductory mathematical
chapter in the {\it Treatise}, Atwood never uses absolute
equations, so that any physical law is always expressed as a
proportion. Thus, in the present case, the factor $1/2$ does not
appear, and the final result is rather written as $s/s^\prime = a
t^2 / a^\prime t^{\prime 2}$. The same applies below, though we
will use modern notations for simplicity of explanation.}

Next, Atwood considers experiments suitable for obtaining the law
for the velocity, whose measurement with the machine is
particularly intriguing (and, mainly, {\it feasible}). Indeed, let
us suppose with Atwood that the instantaneous velocity of the
descending body is requested when it passes at a certain height.
Then a ring is placed at that height on the column with the ruler,
whose ring is designed to remove the additional, unbalanced mass
(whose length exceeds ring's diameter) while allowing the passage
of the main body upon which it is affixed. In such a way, and from
now on, the pulley is completely balanced, and the two bodies
continue to move with constant velocity, whose value is easily
obtained from the ratio of the distance traveled in a given time.
With this trick, the law for the velocity can be tested
experimentally even when a constant force is applied to the body,
and the first result to be obtained is the proportionality between
the instant velocity and the time elapsed: $v \propto t$.
\begin{quote}
During the different times 1 second, 2 second, 3 seconds, etc. the
velocities generated will be those of 6 inches, 12 inches, and 18
inches in a second respectively, being in the same proportion with
the times wherein the given force acts (\cite{Treatise}, p. 324).
\end{quote}
Then:
\begin{quote}
It appears therefore, that if different forces accelerate the same
body from quiescence during a given time, the velocities generated
will be in the same proportion with these forces (\cite{Treatise},
p. 326)
\end{quote}
(again, ``forces'' means ``accelerating forces'' or, simply,
acceleration), that is to say, $v \propto a$ in a constant time.
Finally, as above, the complete law $v = a t$ is obtained by means
of proportions:
\begin{quote}
if bodies be acted in by accelerating forces, which are in the
proportion of 3:4, and for times, which are as 1:2, the velocities
acquired will be in the ratio of $1 \times 3$ to $2 \times 4$, or
as 3 to 8 (\cite{Treatise}, p. 326).
\end{quote}
The story, however, does not end here, as could be hastily
expected. Indeed, other two experiments are described to show
that, for bodies accelerated through the same space $s$, $v
\propto \sqrt{a}$ (\cite{Treatise}, p. 327) and, conversely, under
the action of the same acceleration $a$, $v \propto \sqrt{s}$
(\cite{Treatise}, p. 328). That is to say, in modern terms, $v =
\sqrt{2 a s}$.

Finally, and more importantly, ``the moving force must be in the
same ratio as the quantities of matter moved,'' which is Newton's
second law of dynamics (\cite{Treatise}, p. 328):
\begin{quote}
Experiments to illustrate this truth may be comprised in the
subjoined table. \\
\begin{center}
\begin{tabular}{|c|c|c|c|c|}
\hline Moving & Quantities & Accelerating & Spaces & Velocities \\
forces. & of matter & forces. & described & acquired in \\
 & moved. & & in inches. & inches in a \\
 & & & & second. \\ \hline
 & & & & \\
                         $m$ & $64m$ & $\displaystyle \frac{1}{64}$ & 12 & 12 \\
 & & & & \\
$\displaystyle \frac{3m}{2}$ & $96m$ & $\displaystyle \frac{1}{64}$ & 12 & 12 \\
 & & & & \\
$\displaystyle \frac{3m}{4}$ & $48m$ & $\displaystyle \frac{1}{64}$ & 12 & 12 \\
 & & & & \\
\hline
\end{tabular} %
\end{center}
\end{quote}
${}$ \\
From the first three columns it is evident that the ``moving
force'' is equal to the product of the ``quantity of matter
moved'' times the ``accelerating force'', or, in modern notations,
$F = m a$. It is interesting to compare such a result, directly
obtained from {\it experiments}, with the original {\it Newton's}
formulation of the second law in terms of momentum variation
rather than acceleration (see the discussion in the first of Ref.
\cite{NewtonSuccess}).

The accuracy of the conclusions reached by Atwood obviously depends
on the sensitivity of his machine and on possible systematic
errors primarily related to friction and to the promptness of the
experimenter for time measurements. This last point was considered
only in broad outline by Atwood, who just warned possible other
experimenters to train with the ``simultaneous'' activation of the
pendulum clock with the start of the descending body. This is
evidently - a priori - the major source of inaccuracy, and we will
back on it in the next section. The friction issue, including air
resistance, has been already dealt with above, but here we mention
the fact that Atwood did not limit himself to simply state that
``the effects of friction are almost wholly removed by the
friction wheels'' (\cite{Treatise}, p. 316). Indeed, he quantified
such assertion by making a preliminary experiment:
\begin{quote}
If the weights $A$ and $B$ be balanced in perfect equilibrio, and
the whole mass consists of $63 m$, according to the example
already described, a weight of $1 \frac{1}{2}$ grains, or at most
$2$ grains, being added either to $A$ or $B$, will communicate
motion to the whole, which shews that the effects of friction will
not  be so great as a weight of $1 \frac{1}{2}$ or $2$ grains
(\cite{Treatise}, p. 316).
\end{quote}
Note that a mass of $1m$ corresponded to $\frac{1}{4}$ oz (and
thus $63 m \approx 16$ oz), while 480 grains accounted for 10 oz
(and thus 2 grains $\simeq 1/240$ oz), so that the sensitivity of
the pulley was extremely high. With his device, Atwood was then
able to measure accelerations as low as $1/64$ of the free fall
value (\cite{Treatise}, p. 305), an unprecedented accuracy in such
studies.

This fact was also the reason for the subsequent fortune of
Atwwod's machine.

\subsection{Other topics in the \textit{\textbf{Treatise}}}

The description of the machine and the discussion of the
experiments performed with it occupy only about $10\%$ of Atwood's
{\it Treatise}, so that we cannot leave the issue without even a
rapid mention of the remaining topics covered in the book, again
aimed at establishing the Newtonian paradigm. The major part of
them is devoted to display the theory regarding rectilinear
motion, mainly related to the experiments later considered, with
few intriguing exceptions.

One of these, which is worth to mention, is the discussion of the
resistance opposed to spherical bodies when impinging on given
substances. According to Atwood, the resistance force $F_R$
depends, of course, on the substance considered, this effect being
parameterized by the penetration depth $\delta$, and is
proportional to the square of the diameter $D$ of the body; in
modern notation: $F_R \propto \delta \, D^2$ (\cite{Treatise}, p.
40).

The only other ``exceptions'' we here point out regard, instead,
the affair related to the concepts of momentum and {\it vis viva}.
Although Atwood discusses at length such issues in Sect. IX of his
{\it Treatise} (\cite{Treatise}, p. 356 ff.), he introduces the
relevant quantities well before. The momentum $q = m v$ is
introduced in the discussion of (a particular case of) what we now
call the theorem of momentum:
\begin{quote}
The moving force which communicates, and the force of resistance
which destroys the motion of bodies in the same time, will be in a
compound ratio of the quantities of matter in the moving bodies,
and velocities generated or destroyed (\cite{Treatise}, p. 36).
\end{quote}
In modern notation, for given time interval, $F \propto m v$.
Analogously, the {\it vis viva} $m v^2$ is introduced in the
discussion of (a particular case of) what we now call the theorem
of kinetic energy:
\begin{quote}
If bodies unequal in quantities of matter, be impelled from rest
through equal spaces, by the action of moving forces which are
constant, these forces are in a duplicate ratio of the last
acquired velocities, and the ratio of the quantities of matter
jointly (\cite{Treatise}, p. 29)
\end{quote}
That is, in modern notation, for a constant force acting for a
given distance, $F \propto m v^2$. Echoes of the not yet
extinguished polemic with the Leibnizian view of mechanics in
terms of living forces are present in Sect. IX of the {\it
Treatise}, where Atwood considers the ``conservatio motus''
introduced by Daniel Bernoulli in discussing (one-dimensional)
impact of bodies. While we cannot enter here in such an issue, and
thus refer the interested reader to the existing literature (see,
for example, \cite{Schaffer} and references therein), we only
mention how Atwood posed the physical question. How define the
momentum $q$ of a body in motion? In terms of $m v$ (Atwood, from
Newton) or, rather, $m v^2$ (Bernoulli, from Leibniz)? Atwood's
answer was intimately related to the physical properties of
(one-dimensional) impact of bodies: according to him, Bernoulli's
{\it conservatio motus} necessarily implied $q = m v$.

The only other polemic present in the {\it Treatise} is with J.
Smeaton about the role of friction. While the natural philosopher
Atwood was trying to design frictionless machines, the engineer
Smeaton set out to measure friction's effects. The different
attitudes are evident: while friction is seen by Atwood as an
obstacle to the clear manifestation of the perfect Newtonian laws
of Nature, it is instead an important piece of consideration for
engineers and engine makers. Additional, interesting details on
this polemic, here only trivialized, may be found in
\cite{Schaffer}, where a special emphasis is given on what dubbed
as social function of friction.

Finally, few words should be said on the second great topic of the
{\it Treatise}, that is rotation of bodies. Indeed, Atwood's book
is as well a splendid example of XVIII century's treatise on the
mechanics of rigid bodies, with a number of interesting examples
and applications, an idea of which may be formed just by looking
at the drawings at the end of the book. A clear definition for the
centre of gravity is given, together with a discussion of its
relevance for rigid body's motion. Also, the concept of moment of
inertia is already present in Atwood's {\it Treatise}, implicitly
introduced in the cardinal law for the rotation.
\begin{quote}
In any revolving system, the force which accelerates the point to
which the moving force is applied, is that part of the
acceleration of gravity which is expressed by a fraction, of which
the numerator is the square of the distance at which the force is
applied from the axis, multiplied into the moving force, and the
denominator the sum of all the products formed by multiplying each
particle of the system into the square of its distance from the
axis of motion (\cite{Treatise}, p. 345).
\end{quote}
In formulae, $a = r^2 F / I$ ($I$ being the moment of inertia),
which is just a particular case of the law $M = I \dot{\omega}$
for the torque $M$, with $a = \dot{\omega} r$.

Other key, general results for the rotation of rigid bodies are
considered and discussed in the {\it Treatise}, but their punctual
reconstruction is here unnecessary: Atwood's main aim of the
general achievement of the grandiose Newtonian paradigm has been,
probably, well illustrated from what discussed above.

\

\

\section{The spreading of Atwood's machine: \\ G.S. Poli and J.H. de Magellan}
\label{s4}

\noi The direct spreading of Atwood's machine outside the borders of England was due mainly to the work of two scholars, who were eyewitnesses of the experiments performed by Atwood with his novel machine: namely, the Portuguese J.H. de Magellan and the less known Italian G.S. Poli.

\subsection{A Newtonianist in Naples}

\noi Giuseppe Saverio Poli was born in Molfetta, Italy where he completed his first studies at the Jesuit College which, together with the Convent of the Dominicans, culturally animated this small agricultural and fishing town \cite{PoliBio}. In 1764 his family arranged for him to study at the University of Padua, where 
he graduated in Medicine, and in 1771 moved to Naples practicing as a doctor at the {\it Ospedale degli Incurabili} and, one year later, he was entrusted with the teaching of Geography and History at the ``Nunziatella'' Military Academy. It was during this period that Poli resumed his studies (began in Padua) on electricity and, in 1772-3 he published two books on the formation of and effects induced by thunder and lightning \cite{Polithunder}. In the second of these books, Poli described a series of experiments conducted by himself and questioning about one of the main theories of Benjamin Franklin, i.e. about the adiathermanous property of glass (Franklin was the first, in 1751, to suppose that glass is completely impermeable to heat rays). Poli did not know an explanation for some observed anomalies of electrized glass, but noted that they were not justified by Franklin's dominant theory. Nevertheless, to explain with his model also the insulating effect that the glass sometimes displayed, Poli resorted to the existence of two forms of electricity, which he called ``by source and by contact'' \cite{SchetFrank}. 

As a result of his success within the scientific community in Naples, in 1775 Poli was entrusted with the important task of traveling on behalf of the King of Naples, in order to acquire scientific instruments for the Military Academy and also to study the cultural institutions of the major European capitals. This ``scientific'' journey lasted five years, but very few details are known about it \cite{PoliToscano}; some information about his stay in Cambridge and the meeting with Atwood may be, however, deduced from his later textbook \cite{PoliElementi}.

Meanwhile, spurred by a group of reforming intellectuals, in 1778-9 Ferdinand IV of Bourbon founded in Naples two important institutions: the Royal Academy of Sciences and Humanities, and the Medical School of the {\it Ospedale degli Incurabili} \cite{Borrelli}. Director of the School was Giovanni Vivenzio, who urged the authorities in order to establish in it the teaching of Experimental Physics, with the related ``theater'' where experiments could be performed according to the Newtonian address. The intention of those reformers was to create a new kind of doctors, updated on the latest scientific results, and Vivenzio was able to obtain that the examinations of Anatomy and Experimental Physics were mandatory, and that students could not graduate without a good knowledge on these subjects.

The teaching of Experimental Physics was entrusted to Poli, who was informed in late 1780, when still in England: this was an opportunity to ask the famous instrument maker Jesse Ramsden \cite{Ramsden}, encountered in his stay in England, to build the new machine invented by Atwood in order to study the physical laws of kinematics and dynamics.

\subsection{The Atwood's Machine arrives in the Continent}

\noi According to Poli, Atwood's machine he commissioned to Ramsden would have been useful in his teaching to the young students of the Medical School, as clearly highlighted in the inaugural lecture which he held for the start of the course in Experimental Physics \cite{PoliProlus}. Such inaugural lecture was printed by the Royal Printing Office ({\it Stamperia Reale}), this showing the importance given to that course. The same intention may be deduced from another inaugural lecture, printed in the subsequent year, that for the course in Experimental Physics for the cadets of the ``Nunziatella'' Military Academy \cite{PoliRagion}: here Poli laid the foundations for how he dealt with the study of Nature.

In the meantime, Poli had left England and, in this same year, he gave to the press his textbook on the {\it Elementi di Fisica Sperimentale} \cite{PoliElementi}, already mentioned above. The particular relevance of this first edition\footnote{The first Neapolitan edition of 1781 is rarely mentioned in the literature, while the second Neapolitan edition of 1787 is often quoted as the first one.} of the {\it Elementi} follows from the fact that for the first time a plate appears in it, engraved by de Grado, with the illustration of Atwood's machine (note that Atwood's {\it Treatise} \cite{Treatise} was published only three years later, in 1784). The instrument illustrated in this textbook was the copy ordered to and built by Ramsden; its story, along with those of the other first models of Atwood's machine is quite intruiging.

\begin{figure}
\includegraphics[width=60mm]{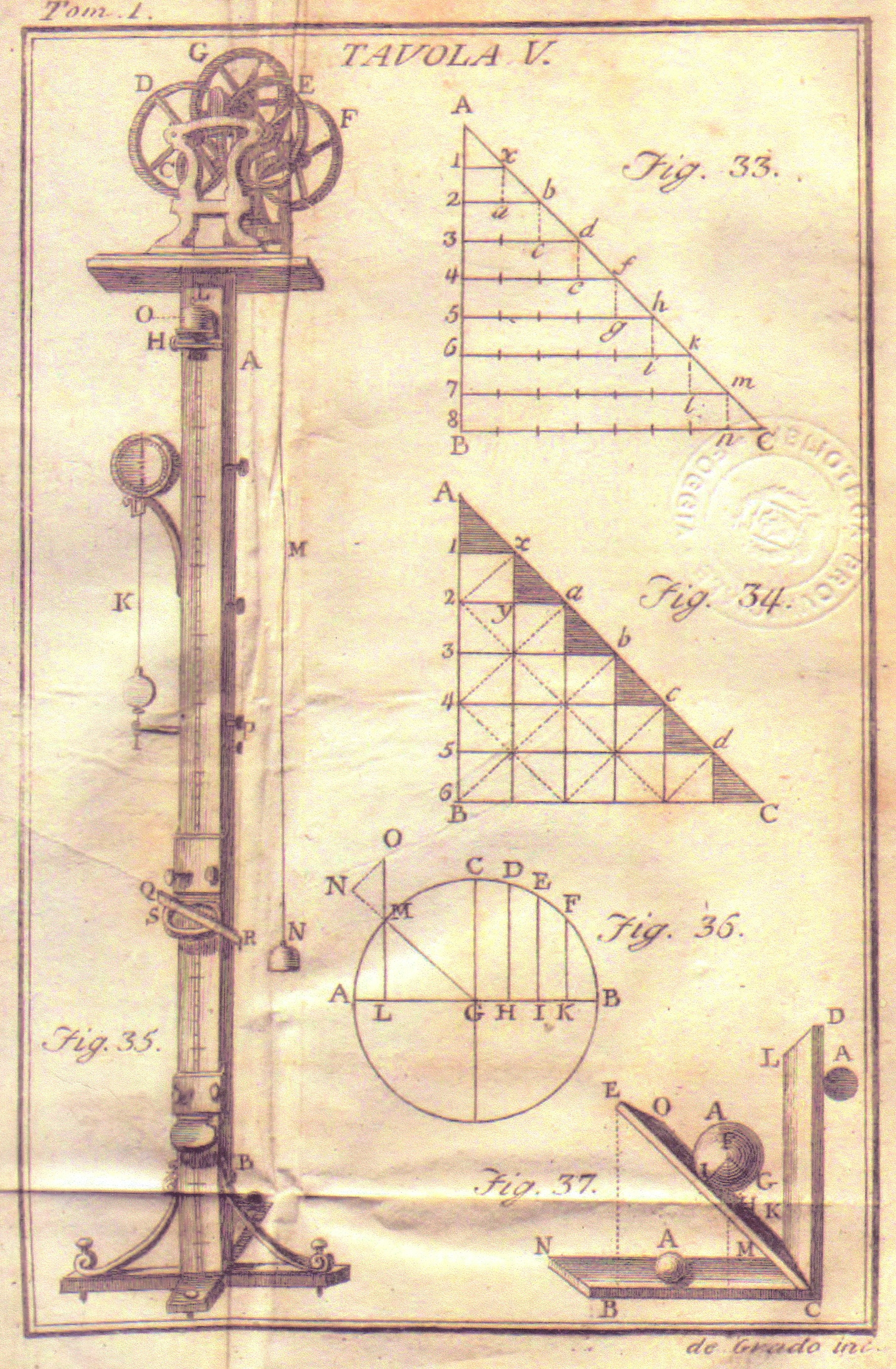}
\caption{The plate in the first edition of Poli's {\it Elementi} \cite{PoliElementi} with the illustration of Atwood's machine built by Ramsden.}
\label{mapoli}
\end{figure}

The copy purchased by Poli for his course at the Medical School and at the Military Academy in Naples arrived in Marseille in October 1782 and was delivered to the countess Zuccheri Stella, who took charge of sending it to Naples so that it could be delivered to Poli:
\begin{quote}
Dal Sig. Gioacchino Bettalli di Parigi mi fu spedito una cassa contenente istrumenti fisici, e pezzi dell'istoria naturale, la quale lui mi dice avere avuto ordine dal Sig. Cav.  Poli di recapitarmela; questa essendomi alla fine giunta, non sapendo io se il d. Sig. Cav. Poli si trova cost\'i, o altrove; e dall'altra parte sapendo per relazione del mio cugino l'Abbate Boscovich, che la d.a cassa deve servire per uso di codesta Universit\`a, ho stimato mio dovere spedirla alla direzione di V.S., sicura, che dar\`a ordini, che sia essa consegnata a chi compete  \cite{Zuccheri}.\footnote{``Mr Gioacchino Bettalli in Paris sent out to me a case containing physics instruments and several specimens of natural history.  He writes that he was commissioned by Mr. Poli to send me that case. Having arrived such a case, since I do not know if Mr. Poli is there or somewhere else, while knowing from my cousin, the Abb\'e Boscovich, that the said case shall serve for use in this University, I have thought it my duty to send it to your lordship, sure that you will give orders that it be given to who be due.''}
\end{quote}
Some months before, another Atwood's machine, built by Adams, arrived in the port of Genoa: it was delivered to Alessandro Volta in May 1781:
\begin{quote}
Al mio arrivo qui ho trovato le sei Casse di Macchine provenienti da Londra, ch'io gi\'a sapeva da qualche tempo essere giunte in Genova [...] Credo non sar\`a discaro a V.E. ch'io Le faccia particolarmente conoscere l'indicata Macchina del Sig. Atwood, la quale riunisce i due pi\'u grandi pregi di facilit\'a ed esattezza nelle sperienze che si fanno con essa \cite{Firmian}.\footnote{``On my arrival I found the six boxes with the machines coming from London, about which I already knew since some time to be arrived in Genoa [...] I think it will not be displeasing to Your Excellency that I make known to You the aforesaid machine of Mr. Atwood, which combines the two greatest advantages of ease and accuracy in the experiences that we can do with it.''}
\end{quote}
These are the only two Atwood's machine that reached Italy in 1781-2. While Poli had been in Cambridge to see Atwood demonstrating his machine, thus deciding to order a copy of it for his course in Naples, Volta didn't. It was the portuguese Jo\~{a}o Jacinto  de Magalh\~{a}es, who met Atwood in England, to write to Volta, then teaching at Pavia, and inviting him to subscribe to Atwood's forthcoming {\it Course of Lectures} \cite{MagVolta1779} and, later, to one of the four prototypes of Atwood's machine built between 1780 and 1782, to be addressed to the Physics Laboratory of the University of Pavia \cite{MagVolta1780}. 

Better known to the English-speaking world by the name under which he published most of his works, Jean-Hyacinthe de Magellan \cite{MagellanBio} was born in Portugal in 1722 and, at the age of eleven, went to an Augustinian monastery in Coimbra where he spent the next twenty years living and studying. The scientific tradition among the Coimbra Augustinians (who knew and studied the works of Newton) allowed him to become familiar with science, particularly astronomy, and after received permission from Pope Benedict XIV to leave the order, in 1755 Magellan started a ``philosophical tour" through Europe, finally settling in 1764 in England. Although he produced no particularly relevant scientific works (the large part of which related to scientific instruments), he is known chiefly for his wide circle of acquaintances and for acting as an intermediary in disseminating new information, mainly about chemistry and experimental physics. Among the people he met during his life, we mention only very few names, including A. Lavoisier and J.J. Rousseau in France, Atwood and J. Priestley in England, Volta and many others. His work and notoriety earned him membership in the Royal Society of London (1774), the Academia das Ci\^encias of Lisbon (1780) and the American Philosophical Society of Philadelphia (1784), as well as corresponding membership in the science academies in Paris, Madrid, Bruxelles and St. Petersburg.

Magellan published the letter quoted above to Volta in a pamphlet of 1780 \cite{Magellan}, where he described the new Atwood's machine for studying dynamics, four years before Atwood himself succeeded in publishing his {\it Treatise} describing all the properties of his device.
\begin{quote}
Cette machine, dans son \'etat a\'ctuel, rend sensible les loix du mouvement {\it uniform\'ement acceler\'e}, ou {\it retard\'e}, de m\'eme que celles du mouvement {\it uniforme}, sans employer qu'un espace moindre de {\it cinq pieds} \& {\it demi}; ce qui la rend extremement commode \& tr\`es avantageuse dans un Cours de Physique. La simplicit\'e \& l'exactitude avec laquelle cette machine rend ce genre d'exp\'eriences \`a la port\'ee des sens, sont encore son plus grande m\'erite: car vous savez que les observations sur la {\it chute des corps}, \& {\it l'acc\'el\'eration de leurs vitesses}, demandent des op\'erations tr\`es d\'elicates, fort difficiles, \& assez laborieuses: \& ce qui plus est, tout-\`a-fait impraticables dans un Cours regulier de Physique Exp\'erimentale \cite{Magellan}.\footnote{``This machine, in its present form, allows the demonstration of the laws of the {\it uniformly accelerated} or {\it retarded} motion, as well as those of {\it uniform} motion, by using a space less than {\it five and a half feet}; making it extremely useful and profitable within a Course of Physics. The simplicity and accuracy with which this machine makes this kind of experiments within the reach of the senses, remain his greatest merit, since you know that the observations on {\it falling bodies} and their {\it accelerations} require very delicate, difficult and laborious operations; and, what is more, all of them is impracticable in a regular course of Experimental Physics.''}
\end{quote}
In the {\it additions et corrections} to his pamphlet, Magellan stated explicitly that there were four machines that were being built:
\begin{quote}
Tandis qu'on imprimoit cette Lettre, j'eus occasion de faire les observations et remarques suivantes, on examinant, comme je vous l'ai promis, Monsieur, la machine qui vous est destin\'ee; \& une autre que j'envoyerai aussit\^ot \`a mon ancien Confrere, les tr\`es-R.P.D. Joachim de l'Assumption, Chanoine R\'egulier d'un m\'erite sort distingu\'e, actuellement Professeur de Physique dans le Monastere Royal de Chanoines R\'eguliers Lateranenses de S. Augustin \`a Mafra, pr\'es Lisbonne. Ces deux machines sont marqu\'ees avec les N. 3 \& 4; parcequ'en effet, on n'a pas encore fait plus, que deux autres machines de cette espece jusqu'\`a present, m\'eme en y comprenant celle de l'inventeur \cite{Magellan}.\footnote{``While printing this Letter, I had occasion to make the following comments and observations, when examining the machine that is for you, Sir, as I have promised to you, and another one that I have to send as soon as possible to my former colleague, the very R.H. Don Joachim de l'Assumption, a very distinguished Canon Regular, currently Professor of Physics in the Royal Monastery of Lateran Canons Regular of S. Augustine in Mafra, near Lisbon. These two machines are marked with N. 3 and 4; indeed, we have not yet built more than two other machines of this kind until now, including also that of the inventor.''}
\end{quote}
The same news (evidently taken by Magellan) is reported in the letter by Volta mentioned above: 
``la Macchina che mi \'e giunta \'e la terza appena che sia stata fatta, non essendosene fino ad ora fabbricate pi\'u che due altre, compresavi quella dell'illustre inventore'' \cite{Firmian}.\footnote{``The machine that has come to me is just the third to have been made, no more than two others having been produced, including that of the illustrious inventor.''} 

Contrary to what sometimes appeared in literature \cite{mala}, it is thus evident that the first copy of Atwood's machine - that ``marked'' with N. 2 - was that realized by Ramsden on behalf of Poli.

\subsection{Intriguing changes}

\noi As recalled above, the major source of errors in the experiments performed with Atwood's machine came from the possible non simultaneous activation of the pendulum clock with the start of the descending body, such ability being left to the promptness of the experimenter. Poli apparently remedied to such possible inconvenience: indeed, the machine manufactured by Ramsden had a {\it novelty} compared to those built by Adams, that is a lever that allowed the experimenter just to start simultaneously the motion of the pendulum and the fall of the mass:
\begin{quote}
Le parti principali di cui \'e composta questa nuova Macchina, sono l'asta verticale AB dell'altezza di cinque piedi e mezzo, divisa in 64 pollici; le cinque picciole ruote C,D,E,F,G; i tre sostegni H,I,S, il pendolo K, ed i pesi convenienti. [...] Come in fatti lo stesso Autore mi ha dimostrato, che anche nella massima velocit\'a, che si suol dare al peso O, la resistenza, che l'aria fa su di esso, a mala pena supera quella del peso di un grano. [...] V'\'e nell'asta AB una molla, guarnita di un bottoncino P, merc\'e la cui pressione si fa si, che nel medesimo istante cadano i due sostegni H ed I, e quindi che il Pendolo K ch'era arrestato dal sostegno I, incomincia ad oscillare nel punto stesso, che il grave O incomincia a discendere. A codesto picciolo meccanismo si \'e data la massima perfezione dal felicissimo genio del celebre Ramsden \cite{PoliElementi}.\footnote{``The main parts that make up this new machine are: the vertical arm $AB$ with an height of five feet and a half, divided into 64-inches; five tiny wheels $C, D, E, F, G, H$; three holders $H, I, S$; the pendulum $K$; and suitable weights. [...] In fact, as the author himself has shown me, even in the case of the maximum speed usually reached by the weight $O$, the resistance offered by the air on it barely exceeds that of the weight of one grain. [...] In the vertical arm $AB$ there is a spring, provided with a button $P$, through the pressure of which it causes the two holders $H$ and $I$ to open at the same moment, and therefore the pendulum $K$, which was kept idle by the holder $I$, begins to oscillate at the same moment in which the weight $O$ begins to descend. The highest perfection of this small mechanism is due to the happy talent of the renowned Ramsden.''} 
\end{quote}
In Fig. \ref{matwood}, reproduced here from the Atwood's {\it Treatise}, we find the original design of the machine: all the models built by Adams conform to such design. In these early prototypes, there are two vertical arms: one supports the pendulum clock for the measurement of time, while the other arm supports the ruler and the latches for the two weights with the gear of the pulley. The machine realized by Ramsden, instead, has a different making: as it is evident from the plate engraved by de Grado for the 1781 edition of Poli's {\it Elementi} (see Fig. \ref{mapoli}), there is a single vertical arm (parallelepiped-shaped) supporting both the pendulum clock and the latches for the two weigths. The trigger mentioned above is as well anchored to the single vertical arm, allowing an easy start of the experiments. Convinced that such a trigger would have reduced the measurement errors, Poli directly spoke of this improvement with Atwood who, however, did not appear entirely persuaded of its effectiveness, or, rather, was convinced that it would not have improved the measurements. Nevertheless, as a matter of fact, Atwood mentioned this problem - though giving only a broad outline of it - in his {\it Treatise}, this being probably a reminiscence of Poli's advice (\cite{Treatise}, p.308). Interestingly enough, the machines that were built later (during the XIX century) had all this trigger, though the instrument makers chose to use two vertical arms instead of only one, by keeping separate the pendulum clock from the weigths. 

The set of five friction wheels was quite immediately simplified: in the 1801 edition of the {\it Trait\'e \'el\'ementaire de Physique} by A. Libes \cite{Libes}, for example, Atwood's machine was already represented with just a simple pulley on the top of a vertical arm with a ruler. Even without resorting to such extreme simplifications, it is nevertheless a matter of fact that, in any of the copies realized in the XIX century, the removable apparatus with the friction wheels in the original prototype was replaced by a fixed one: the original intent of a ``universal'' machine for translational and rotational motions disappeared.

An apparent exception to such boost to simplify is given by Poli's {\it Elementi} \cite{PoliElementi} which, in any of its 23 editions (ranging from 1781 to 1837, i.e. well beyond the death of Poli), Atwood's machine is invariably depicted as in the 1781 edition,\footnote{An irrelevant difference appearing since the second edition concerns only the initial position of the pendulum: in the 1787 edition, it is sketched at the starting of the oscillation, rather than at rest.} this being evidently related to the original instrument that Poli had at his disposal. It is interesting to note that, although Poli himself gave a simplified presentation in his treatise of Atwood's machine and of the experiments that could be performed with it, as any other author did, nevertheless he stressed his will to revisit the subject in a later work:
\begin{quote}
Tutte le rapportate dottrine riguardanti la discesa de' gravi, rintracciate mirabilmente dall'immortal Galileo, render si possono sensibilissime, ed evidenti, merc\'e di una Macchina inventata, non \'e molti anni, dal Signor Atwood, Professore di Fisica nell'Universit\`a di Cambridge, e mio rispettabile Collega nella Societ\`a R. di Londra. Converrebbe scrivere un intiero trattato per dare una compiuta idea di siffatta Macchina, e per indicare la maniera, onde si debbono con essa istituire tutti gli esperimenti. Sar\`a questo in qualche parte il soggetto di un'altra mia Opera \cite{PoliElementi}.\footnote{``Any result concerning the falling of bodies, as admirably deduced by the immortal Galileo, can be made extremely accurate and evident by means of a machine invented, not many years ago, by Mr. Atwood, Professor of Physics at the University of Cambridge and my respected colleague at the Royal Society of London. It would be appropriate to write an entire treatise in order to give an idea of such a machine and to describe how any experiment could be performed with it. This will be somewhere the subject of another work of mine.''} 
\end{quote} 
However, not only Poli did not add a specific chapter on this subject in any of the subsequent editions of his {\it Elementi}, but he did not publish even any other work regarding Physics, devoting himself in those years to the writing of a major work concering testaceans \cite{PoliBio}.

\subsection{A different use}

\noi The small (but relevant) changes introduced in the realization of copies of Atwood's machine seen above were, of course, functional to a better operation of the machine, but the ``simplification'' of it as a whole intervened already at the beginning of the XIX century does not call for a similar explanation. This can be searched, instead, by looking at the use that scholars made of the machine which, as already envisaged in the quotation above from Poli's {\it Elementi}, was substantially {\it different} from the original Atwood's intention. Even more explicitly than Poli, the following words by Volta are illuminating:
\begin{quote}
V.E. pu\'o giudicare di qui se [la Macchina] \'e novissima: lo \'e tanto, che non \'e comparsa ancora l'Opera che il Sig. Atwood medesimo promette di pubblicare sopra questa sua Macchina di Dinamica, dove la descrizione ne sar\`a pi\'u compiuta di quella che or ci da il Sig. Magellan [...] Ho ripetute io gi\'a le principali sperienze proposteci in essa ne' 14 problemi, e le ho variate in pi\'u maniere; e sempre l'esito ha corrisposto alla teoria con una precisione, che maggiore desiderar non si potrebbe. Le leggi della caduta dei gravi son messe cos\'i chiaramente e distintamente sott'occhio, che anche chi nulla conoscesse della teoria, vi \'e tosto condotto e le intende a maraviglia. Da qualche giorno che ho messo alla prova la Macchina non so quasi occuparmi d'altro, tanta \'e la soddisfazione che ne ritraggo \cite{Firmian}.\footnote{``Your Excellency will judge from this whether [the machine] is new: it is so much new that has not yet appeared the work that Mr. Atwood himself has promised to publish about his dynamical machine, where it will be given a more complete description of what we now have from Mr. Magellan [...] I have already repeated the key experiments proposed by him in the reported 14 problems, having changed them in several ways; the results always conformed to the theory with an accuracy, that you could not desire more. The laws on the falling of bodies are made so clear and distinct, that even those who know nothing of the theory are led to fully understand them. I tested the machine since few days, and I almost cannot take care of other things, 
so great is the satisfaction I feel with it.''}
\end{quote}
What was the original motivation for the construction of the machine - that of displaying the Newtonian paradigm - has now been changed: its use was quite soon limited just to perform several illustrative
experiments on the falling of bodies with a single and accurate device.
This is, indeed, the use made of Atwood's machine since the end of the XVIII century until recent times (whereas, sometimes, the experimental study on the falling of bodies is replaced by generic studies on uniform and uniformly accelerated motion).

The reasons for this apparent change of mind are, again, contained in the quotation above: the machine arrived in the Continent well before the appearance of the {\it Treatise} where Atwood explained its original use, while the description of the experiments that could be performed with it was spread only through the work of the eyewitnesses Poli and, especially, Magellan. While, on the one hand, this implied a rapid fortune of Atwood's machine, given the success of the Poli's {\it Elementi} with its 23 editions and the indefatigable work as intermediary of Magellan, on the other hand it evidently allowed the subsequent scholars to make (only) a different use of the machine, given the already occurred achievement of the Newtonian paradigm. Furthermore, Atwood himself contributed indirectly to such direction, since he never reissued the 1784 edition of his {\it Treatise}, having later changed his interests, as recalled above.

\

\

\section{Summary}

\noi The structure of Newtonian physics is, as well known, based on the organization of scientific knowledge as a series of mathematical laws and, according to early codification by Galilei, such laws requires experimental validation. In the XVIII century, physical demonstrations took place in different ways and for different audiences, ranging from academic courses to popular lectures. In the present paper, we have shown how the Newtonian paradigm was definitively accepted in science courses - in England as well as in the Continent - by means of the dynamical machine invented by Atwood in late 1770s just for this purpose. Although aware that the ultimate test of Newton's mechanics would have come from experiments showing the effect of variable forces, the experimental conditions of his epoch forced Atwood to turn to constant forces, for which he designed a single machine in order to test, in simple experiments, {\it all} the kinematical and dynamical laws for those forces, as coming out from Newton's mechanics. Particularly relevant is the mechanism he devised to measure the velocity acquired by the body during its accelerated motion: in Galilei's inspired experiments on the free fall or on the motion along an inclined plane, indeed, only the proportionality between the traveled spaces and the square of the elapsed time could be established. But, probably, the astonishing result was the unprecedented accuracy with which Atwood tested the Newtonian laws of motion, being able to measure acceleration as low as $1/64$ of the free fall value. As described in his {\it Treatise} of 1784, Atwood's original aim was not limited to the mechanical laws of the rectilinear motion, the machine having to serve also for studying the rotation of bodies (in the {\it Treatise}, a number of experiments are described concerning this topic, with the aid of additional parts, never effectively realized), but such an incredible accuracy catalyzed the interest of any of the subsequent scholars who used the machine in their demonstrations. 

The spreading of Atwood's machine outside England occurred well before the appearance of the {\it Treatise}, where it was described along with the experiments to be performed with it. In fact, some scholars who had the opportunity to attend Atwood's demonstrations in Cambridge in the late 1770s, realized immediately the importance of the novel machine, and disseminate the news, even subscribing (or suggesting the subscription to other scholars) to the acquisition of copies of that machine. Thus, it was the Portuguese Magellan the first to publish (in 1780) a pamphlet where the new machine was broadly described, along with a set of experiments concerning uniform and accelerated motion, in the form of a letter addressed to Volta in Pavia. Instead, it was the Italian Poli to report (in 1781) for the first time an illustration of the novel machine, realized on the copy ordered to the instrument maker Ramsden, in his textbook where a choice of experiments are described as well. 

The model manufactured by Ramsden (the second one ever realized, including the original one owned by Atwood) introduced an additional device, suggested by Poli, in order to trigger the simultaneous activation of the pendulum clock and the start of the descending mass. Clearly aimed at a better operation of the machine and, consequently, at a reduction of the measurement errors, this additional lever was always included in later copies of the machine during the XIX century, irrespective of the ``simplification'' of Atwood's machine (the removable set of five friction wheels was replaced by a fixed set of wheels or even just a simple pulley) that was going on already at the very end of XVIII century.

Such changes which were occurring on the machine are emblematic of the different use made of it. Once  Newton's mechanics was definitively accepted in academic courses as the only possible theory of motion, Atwood's machine did not serve anymore as a device displaying the success of the Newtonian paradigm. Its use then changed accordingly: several illustrative experiments on the falling of bodies, or even just on uniform or uniformly accelerated motion may be performed with a single and accurate machine. This is, indeed, the use made of Atwood's machine until now.

The historical case studied here, therefore, allows us to recognize the relevant role played by a properly devised instrument in the acceptance of a new paradigm by non-erudite scholars, in addition to the traditional ways followed by erudite ones (almost exclusively considered in the literature), where mathematical, philosophical or even physical reasoning certainly dominates over {\it machine philosophy}.

\




\begin{thebibliography}{99}

\bibitem{soc}
J.E. McClellan III, ``Learned Societies'' in {\it Encyclopedia of the Enlightenment}, edited by A.C. Kors, Oxford University Press, Oxford, 2003. \\
C.C. Gillispie, {\it Science and Polity in France at the end of the Old Regime}, Princeton University Press, Princeton, 1980.

\bibitem{Feingold}
M. Feingold, {\it The Newtonian Moment: Isaac Newton and the Making of Modern Culture},New York Public Library, New York, 2004.

\bibitem{ugo}
J.F. Baillon, Studies in History and Philosophy of Science Part A {\bf 35} (2004) 533-548.
See also G.C. Gibbs, {\it Huguenot contributions to the intellectual life of England, c. 1680 Ð c. 1720, with
some asides on the process of assimilation}, in J.A.H. Bots and G.H.M. Posthumus Meyjes (eds.), {\it La 
r\'evocation de l'Edit de Nantes et les Provinces-Unies 1685}, APA Holland University Press, Amsterdam, 1986.

\bibitem{Stewart}
L. Stewart, {\it The rise of public science: Rhetoric, technology and natural philosophy in Newtonian
Britain, 1660Ð1750}, Cambridge University Press, Cambridge, 1992.

\bibitem{Jacob}
M.C. Jacob, {\it  The Cultural Meaning of the Scientific Revolution}, Temple University Press, Philadelphia, 1988.

\bibitem{Wolff}
R.S. Calinger, Journal of the History of Ideas {\bf 30} (1969) 319-330.

\bibitem{Hanna}
B.T. Hanna, {\it Polini\`ere and the teaching of physics at Paris: 1700-1730}, in P. Gay (ed.), {\it Eighteenth-Century Studies Presented to Arthur M. Wilson}, University Press of New England, Hanover (New Hampshire), 1972.

\bibitem{Schet}
E.Schettino, {\it L'insegnamento della fisica sperimentale a Napoli nella seconda met\'a del Settecento}, Studi Settecenteschi, vol. 18, Bibliopolis, Naples, 2001. 

\bibitem{Principia}
I. Newton, {\it Philosophiae Naturalis Principia Mathematica}, London, 1726 (third edition).

\bibitem{NewtonSuccess}
See, for example, R.S. Westfall, {\it Force in Newton's Physics.
The Science of Dynamics in the Seventeenth Century}, MacDonald,
London, 1971. See also R. McCormmach, {\it Speculative truth:
Henry Cavendish, natural philosophy, and the rise of modern
theoretical scienc}, Oxford University Press, Oxford, 2004.

\bibitem{Schaffer}
S. Schaffer, Osiris {\bf 9} (1994) 157-182.

\bibitem{Desag}
A.R. Hall, ``Desaguiliers, John Theophilus'' in {\it Dictionary of Scientific
Biography}, edited by C.C. Gillispie, vol. V, Charles Scribner's
Sons, New York, 1972.

\bibitem{AtomPowers}
For a thoughtful examination of the studies by Newton on the internal structure of matter, see for example A. Thackray, {\it Atom and Powers. An Essay on Newtonian Matter Theory and the Development of Chemistry}, Harvard University Press, Cambridge (Mass.), 1970. 

\bibitem{DesagCourse}
J.T. Desaguliers, {\it A Course of Experimental Philosophy}, vol. I, London, 1734; vol. II, London, 1744.

\bibitem{Mariotte}
E. Mariotte, {\it A treatise of the motion of water, and other fluids: with the origin of fountains of springs, and the cause of winds... Written originally in French... And translated into English, with several annotations for explaining the doubtful places, by J. T. Desaguliers, M. A.}, London, 1718.

\bibitem{Gravesande}
W.J. 's Gravesande, {\it Mathematical elements of natural philosophy, confirm'd by experiments: or, an introduction to Sir Isaac Newton's philosophy. Written in Latin by the late W. James s'Gravesande, LL.D. ... Translated into English by the late J.T. Desaguliers, LL.D. F.R.S.}, London, 1747.

\bibitem{DesagPoem}
J.T. Desaguliers, {\it The Newtonian system of the world, the best model of government: An allegorical
poem}, London, 1728.

\bibitem{DesagDiss}
J.T. Desaguliers, {\it A dissertation concerning electricity}, London, 1742.

\bibitem{sGrave}
A.R. Hall, ``'s Gravesande, Willem Jacob'' in {\it Dictionary of Scientific
Biography}, edited by C.C. Gillispie, vol. V, Charles Scribner's
Sons, New York, 1972.

\bibitem{Opticks}
I. Newton, {\it Opticks or, a treatise of the reflexions, refractions, inflexions and colours of light}, London 1704 (Latin version, 1706; revised version, 1718).

\bibitem{Muss}
D.J. Struik,  ``Musschenbroek, Petrus van'' in {\it Dictionary of Scientific
Biography}, edited by C.C. Gillispie, vol. IX, Charles Scribner's
Sons, New York, 1979. \\
L. N. Kryzhanovskii, Sov. Phys. Usp. {\bf 34} (1991) 265-268.

\bibitem{Elementa}
P. van Musschenbroek, {\it Elementa Physicae Conscripta in Usus Academicos}, Leyden, 1734.

\bibitem{Insti}
P. van Musschenbroek, {\it Institutiones Physicae Conscriptae in Usus Academicos}, Leyden, 1748.

\bibitem{Optfra}
I. Newton, {\it Trait\'e d'optique sur les reflexions, refractions, inflexions, et couleurs de la lumiere... Traduit de l'Anglois par M. Coste}, Amsterdam 1720.

\bibitem{alllight}
A.R. Hall, {\it All Was Light: An Introduction to Newton's Opticks}, Oxford University Press, Oxford, 1995.

\bibitem{Nollet}
J.L. Heilbron, ``Nollet, Jean-Antoine'' in {\it Dictionary of Scientific
Biography}, edited by C.C. Gillispie, vol. X, Charles Scribner's
Sons, New York, 1981.

\bibitem{Lecons}
J.A. Nollet, {\it Lecons de physique experimentale}, six volumes, Paris, 1743-1748.

\bibitem{Voltaire}
Francois-Marie Arouet, {\it El\'emens de la Philosophie de Neuton, Mis \'a la port\'ee de tout le monde. Par Mr. De Voltaire}, Amsterdam, 1738.

\bibitem{curvature}
Charles-Marie de la Condamine, {\it Journal du voyage fait par ordre du Roi \`a l'\'equateur servant
d'introduction historique \`a la m\'esure des trois pr\'emiers degr\'es du M\'eridien}, Paris,  1751; \\
A.C. Clairaut, {\it Th\'eorie de la figure de la terre, tir\'ee des principes de l'hydrostatique}, Paris, 1743.

\bibitem{Palladino}
A complete transcription of the letter (in Italian) is in F. Palladino, 
Rend. Acc. Naz. Scienze detta dei IX {\bf IX} (1985) 334-336.

\bibitem{Petro}
P. van Musschenbroek, {\it Elementa physicae conscripta in usus academicos: quibus nunc primum in gratiam studiosae juventutis accedunt ab alienis manibus ubique auctaria et notae, disputatio physico-hist	orica de rerum corporearum origine, ac demum de rebus coelestibus tractatus}, Naples, 1745.

\bibitem{Torrini}
About the {\it Disputatio} by Genovesi and on his attempt to oppose the philosophical interferences encountered in Newtonian oriented textbooks, see M. Torrini, {\it La discussione sulla scienza}, in G. Pugliese (ed.), {\it Storia e Civilt\'a della Campania. Il Settecento}, Electa, Naples, 1994.

\bibitem{DellaTorre}
G.M. Della Torre, {\it Scienza della Natura}, two volumes, Naples, 1748-9.

\bibitem{Lalande}
J.J. de Lalande, {\it Voyage d'un Francais en Italie, fait dans les ann\'ees 1765 et 1766}, Paris, 1769.

\bibitem{Borrelli2}
A. Borrelli, {\it Istituzioni scientifiche Medicina e societˆ. Biografia di Domenico Cotugno (1736-1822)}, Olschki, Florence, 2000.

\bibitem{PoliToscano}
M. Toscano, {\it Gli Archivi del Mondo. Antiquaria, storia naturale e collezionismo nel secondo Settecento},
Edifir, Firenze 2009. 

\bibitem{PoliElementi}
G.S. Poli, {\it Elementi di Fisica Sperimentale}, Naples, 1781.

\bibitem{Alumni}
``Atwood, George'' in J. \& J.A. Venn, {\it Alumni
Cantabrigienses}, Cambridge University Press, Cambridge,
1922-1958.

\bibitem{Bios}
E.M. Cole, ``Atwood, George'' in {\it Dictionary of Scientific
Biography}, edited by C.C. Gillispie, vol. I, Charles Scribner's
Sons, New York, 1970.

\bibitem{Description}
G. Atwood, {\it A Description of the Experiments Intended to
Illustrate a Course of Lectures on the Principle of Natural
Philosophy}, London, 1776.

\bibitem{Treatise}
G. Atwood, {\it A Treatise on the Rectilinear Motion and Rotation
of Bodies With a Description of Original Experiments Relative to
the Subject}, Cambridge, 1784.

\bibitem{Analysis}
G. Atwood, {\it An Analysis of a Course of Lectures on the
Principles of Natural Philosophy}, Cambridge, 1784.

\bibitem{Biblio}
See Atwood's bibliography in Ref. \cite{Bios}.

\bibitem{17961798}
G. Atwood, Phil. Trans. {\bf 86} (1796) 46-130; {\it ibid.}, {\bf
88} (1798) 201-310.

\bibitem{arches}
G. Atwood, {\it A Dissertation on the Construction and Properties
of Arches}, London, 1801.

\bibitem{Hall}
F.G. Hall, The Mathematical Gazette {\bf 11} (1922) 108-110.

\bibitem{Sully}
H. Sully, {\it Description Abreg\'ee d'une Horloge d'une Nouvelle
Invention, pour le Juste Mesure du Temps}, Paris, 1726 (pp. 9-10).

\bibitem{PoliBio}
The most recent biography of Poli is in G. De Gennaro, {\it Uno scienziato alla corte dei Borbone di Napoli: Giuseppe Saverio Poli (Molfetta 1746 - Napoli 1825)}, Risorgimento e mezzogiorno {\bf 1-2} (2006) 91-95.

\bibitem{Polithunder}
G. S. Poli, {\it La formazione del tuono, delle folgore e di varie altre meteore giuste le idee del sig Franklin}, Naples, 1772; {\it Riflessioni intorno agli effetti di alcuni fulmini}, Naples, 1773.

\bibitem{SchetFrank}
E. Schettino, {\it Franklinists in Naples in the second half of the 18th century}, in Proceedings of the XX Congresso Nazionale della Societ\`a Italiana degli Storici della Fisica e dell'Astronomia, Naples, June 1-3,  2000, pp. 347-352.

\bibitem{Borrelli}
A.Borrelli, {\it G.S. Poli e la Scuola medica degli Incurabili di Napoli}, Conference at the Universit\`a Popolare di Molfetta, November 2007; A. Borrelli, {\it Le origini della Scuola medica dell'Ospedale degli Incurabili di Napoli}, Archivio storico della Provincia napoletana {\bf CXVIII} (2000) 135-149.

\bibitem{Ramsden}
R.S. Webster, ``Ramsden, Jesse'' in {\it Dictionary of Scientific
Biography}, edited by C.C. Gillispie, vol. XI, Charles Scribner's
Sons, New York, 1986.

\bibitem{PoliProlus}
G.S. Poli, {\it Breve ragionamento intorno all'eccellenza dello studio della natura, e ai seri vantaggi che da quello si possono ritrarre; premesso al corso di Fisica Sperimentale, destinato a farsi nel Regio Ospedale degl'Incurabili, da Giuseppe Saverio Poli, professore di detta scienza nella regia universit\`a}, Stamperia Reale, Naples, 1780.

\bibitem{PoliRagion}
G.S. Poli, {\it Ragionamento intorno allo studio della Natura, composto, e recitato da Giuseppe Saverio Poli nell'Accademia del Battaglione R. Ferdinando, in occorrenza di dover dare ivi un corso di Fisica Sperimentale}, Naples, 1781.

\bibitem{Zuccheri}
Countess Zuccheri Stella to Marquis Della Sambuca, 19 October 1782; reported in \cite{Borrelli}.

\bibitem{Firmian}
A. Volta to Count Firmian, 1 May 1781; in A. Volta, {\it Epistolario}, Zanichelli, Bologna, 1951.

\bibitem{MagVolta1779}
J.H. Magellan to A. Volta, 9 April 1779; in A. Volta, {\it Epistolario}, Zanichelli, Bologna, 1951.

\bibitem{MagVolta1780}
J.H. Magellan to A. Volta, 21 November 1780; in A. Volta, {\it Epistolario}, Zanichelli, Bologna, 1951.

\bibitem{MagellanBio}
S. Pierson, ``Magellan, Jean-Hyacinthe'' in {\it Dictionary of Scientific
Biography}, edited by C.C. Gillispie, vol. IX, Charles Scribner's
Sons, New York, 1979. \\
M. Villas-Boas, {Jo\~{a}o Jacinto  de Magalh\~{a}es - Um empreendedor cientifico na Europa do s\'eculo XVIII}, Aveiro, Fundac\~{a}o Jo\~{a}o Jacinto  de Magalh\~{a}es, 2000.

\bibitem{Magellan}
J.H. Magellan, {\it Description d'une machine nouvelle de dynamique, invent\'ee par Mr. G. Atwood... dans une lettre adress\'ee a Monsieur A. Volta}, London, 1780.

\bibitem{mala}
I.M. Malaquias and M.F. Thomaz, {\it Scientific communication in the XVIIIth century: The case of John Hyacinth de Magellan}, Physis {\bf 31} (1994) 817-834.

\bibitem{Libes}
A. Libes, {\it Trait\'e \'el\'ementaire de Physique, pr\'esent\'e dans un ordre nouveau, d'apr\'es les d\'ecouvertes modernes}, Paris, 1801.

\end{thebibliography}
\end{document}